\documentclass[epj]{webofc}
\usepackage{bm}
\usepackage[T1]{fontenc} 
\usepackage[varg]{txfonts}   
\wocname{QCD@Work 2014}
\woctitle{QCD@Work 2014 }
\begin{document}
\title{Bottomonium in the plasma}
%
%
\subtitle{Lattice results}

\author{
        Gert Aarts\inst{1}\and
        Chris Allton\inst{1}\and
        Wynne Evans\inst{1}\and
        Pietro Giudice\inst{2}\and
        Tim Harris\inst{3}\and
        Aoife Kelly\inst{4}\and \\
        Seyong Kim\inst{5}\and
        Maria Paola Lombardo\inst{6}\and
        Sinead Ryan\inst{3}\and
        Jon-Ivar Skullerud\inst{4}
}

\institute{Department of Physics, College of Science, Swansea University, Swansea SA2 8PP, U.K.
\and
Institut f\"ur Theoretische Physik, Universit{\"a}t M{\"u}nster, D-48149 M{\"u}nster, Germany
\and School of Mathematics, Trinity College, Dublin 2, Ireland
\and Department of Mathematical Physics, Maynooth University, Maynooth, Co.Kildare, Ireland
\and Department of Physics, Sejong University, Seoul 143-747, Korea
\and INFN, Laboratori Nazionali di Frascati, I-00044 Frascati, Italy}

\abstract{We present results on the heavy quarkonium spectrum
and spectral functions obtained by performing large-scale
simulations of QCD for temperatures ranging from about 100 to
500 MeV, in the same range as those explored by LHC 
experiments. We discuss our method and perspectives for 
further improvements towards the goal of full control
over the many systematic uncertainties of these studies.}
\maketitle
\section{Introduction}
\label{intro}
Most ordinary hadrons can only exist up to temperatures of about 
150--170 MeV.
Beyond that, chiral symmetry is restored and confinement is lost.
We know that this hot state of matter --- the Quark-Gluon Plasma (QGP) --- 
existed in the early universe: the transition from the QGP to the 
hadronic world  is the latest cosmological transition. 
The QGP can be re-created in accelerators: the talk by 
Roberta Arnaldi~\cite{Arnaldi} provides an
excellent introduction into the  status of this rich experimental program.

At low temperature the thermal medium consists of a gas
of light pions, while towards infinite temperature
quarks and gluons become free, with a corresponding increase
of the pressure.  After a debate lasting several years
a consensus has been reached on 
how to interpolate between these two different, limiting regimes 
\cite{Ding:2014kva}. It turns out 
that  there is a large intermediate temperature range
 which is not amenable to any analytic approaches, even when
the most sophisticated high temperature expansions and model 
analyses are being used.
This is the region explored by experiments, and this is  where
our lattice simulations are being performed. 

Hadrons are of course dramatically affected in the QGP:
the light quarks
lose their dynamical masses and chiral partners approach degeneracy.
Quark--antiquark states bound by  long-distance, confining forces dissolve. 
It is very remarkable that heavy quarkonia behave very
  differently in this respect, as their fundamental states might well persist into
the plasma: indeed heavy quarks and antiquarks  are bound by short range Coulombic 
interactions which are  not immediately affected by temperatures of the order
of 200 MeV. Experimental evidence has been reviewed 
at this meeting~\cite{Das} and our motivation   
is to provide a solid theoretical baseline for these studies. 

A comprehensive review has recently appeared \cite{Andronic:2014sga}, 
and we concentrate here on our own recent 
work~\cite{Lat,Conf,Aarts:2014cda,Aarts:2013kaa}. 
The next section is an introduction 
into spectral functions and related methodology. 
Then we give an overview of 
 quarkonia in the Quark-Gluon Plasma. 
The following section is devoted to a more detailed presentation of
the bottomonium results. We close with a brief discussion.

\section{Relativistic and non-relativistic spectral functions}
Spectral functions  play an important role in understanding how
elementary excitations are modified in a thermal medium. 
In a relativistic field theory approach the  temperature $T$
is realized through (anti)periodic boundary conditions in the Euclidean time 
direction and the spectral decomposition  of a  zero-momentum  
Euclidean propagator $G(\tau)$ at finite temperature  is given by
\begin{equation}
\label{eq:K}
G(\tau) = \int_{0}^\infty \frac{d\omega}{2\pi}\,  K(\tau,\omega)
\rho(\omega),\hskip 1truecm 0 \le \tau < \frac{1}{T}, 
\end{equation}
where $\rho(\omega)$ is the spectral function and  the kernel $K$ is
given by
\begin{equation}
K(\tau,\omega) =
\frac{\left(e^{-\omega\tau} + e^{-\omega(1/T - \tau)}\right)}
{1 - e^{-\omega/T}}. 
\end{equation}
The $\tau$ dependence of the kernel reflects the periodicity
of the relativistic propagator in imaginary  time, as well
as its $T$ symmetry. The Bose--Einstein distribution, 
intuitively,  describes the wrapping around the periodic box which
becomes increasingly important at higher temperatures. When
the significant $\omega$ range greatly exceeds the temperature,  
$K(\tau,\omega) \simeq \left(e^{-\omega\tau} + e^{-\omega(1/T - \tau)}\right)$:
backwards and forwards propagations are  decoupled and the spectral
relation reduces to 
\begin{equation}
G(\tau) = \int_{\omega_0}^\infty\frac{d\omega'}{2\pi}\, \exp(-\omega'\tau)\rho(\omega').
\end{equation}
This approximation holds true in NRQCD:  
the interesting physics takes place around the two-quark threshold,
$\omega\sim 2M \sim 8$ GeV for $b$ quarks, which is still much
larger than our 
temperatures $T < 0.5$ GeV.  In our  applications, following
ref.\  \cite{Burnier:2007qm}, we will change variable 
 $\omega=2M+\omega'$. 

Turning to the actual computational methodology,  
the calculation of the spectral
functions using Euclidean propagators as an input is a difficult,
ill-defined problem.  We will tackle it using the Maximum Entropy
Method (MEM) \cite{Asakawa:2000tr}, which has proven successful in a
variety of applications.  We have studied the
systematics carefully,
including the dependence on the set of lattice data points in time,
and on the  default model  $m(\omega)$ which enters in the
parametrisation of the spectral function, 
\begin{equation}
\rho(\omega) = m(\omega) \exp \sum_k c_ku_k(\omega),
\end{equation}
where $u_k(\omega)$ are basis functions fixed by the kernel $K(\tau,\omega)$
and the number of time slices, while  the coefficients $c_k$ are to be
determined by the MEM analysis \cite{Asakawa:2000tr}. We find that the
results are insensitive to the choice of default model, provided that
it is a smooth function of $\omega$, and we will provide some 
examples in the next section.  

Recently, an alternative Bayesian reconstruction of the spectral
functions has been proposed in ref. \cite{Rothkopf:2011ef,Burnier:2013nla},
and applied to the analysis of HotQCD configurations \cite{Kim:2014iga}.
Some preliminary results for the bottomonium spectral functions obtained
using this new reconstruction on our ensembles
became available after the QCD@work meeting 
and have been presented at recent conferences \cite{Lat,Conf}.

\section{Overview of our quarkonium results }
The results on bottomonium presented in this note should be framed in the
broader context of studies of quarkonia as QCD thermometers, either from 
lattice first principles simulations, or from a lattice-informed potential
model approach. Our most recent results for bottomonium
\cite{Aarts:2014cda} have been obtained by analysing gauge
field configurations with two active light quarks and one heavier quark.
The lighter quarks are still heavier than the physical up and down  quarks
as  at $T=0$, $m_\pi/m_\rho \simeq 0.4$, while the mass of the heavier quark
is close to the strange mass. These results can be contrasted with earlier
ones obtained with an infinite `strange' mass (two active
  flavours) \cite{Aarts:2011sm,Aarts:2012ka}: in brief summary, we have
found that the results from the different ensembles are broadly consistent,
and we defer a more detailed comparison to future work. 
\begin{figure}
\centering
\includegraphics[width=13cm,clip]{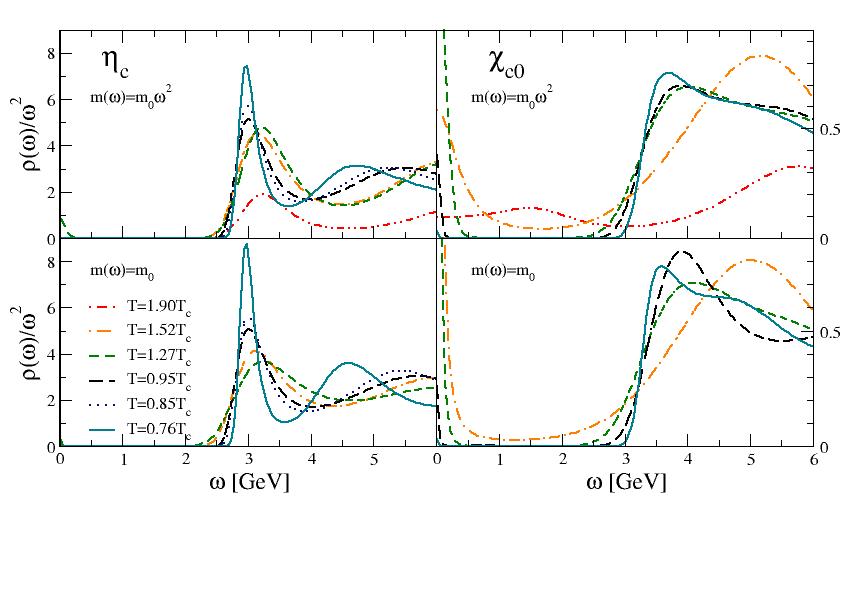}
\vspace{-1.8cm}
\caption{The spectral function for the charmonium states
  $\eta_c$ (S-wave) and $\chi_{c0}$ (P-wave) for varying temperatures, using
  gauge field ensembles with dynamical u, d and s quarks. The results from
two different default models are shown in the upper and lower
diagrams respectively to demonstrate the stability of the analysis.}
\label{fig:charm}   
\end{figure}

The spectral functions of the charmonium states have been studied  
as a function of both temperature and momentum, using as input 
relativistic propagators with two light quarks
\cite{Aarts:2007pk,Kelly:2013cpa} and, more recently, including
  the strange quark. 
These most recent results are shown in fig.~\ref{fig:charm}, for
temperatures ranging between $0.76 T_c$ and $1.9 T_c$. The sequential
dissolution of the peaks corresponding to the S- and P-wave
states is clearly seen. Transport coefficients can be obtained from the
low frequency domain, and this is an important aspect of our
research \cite{Amato:2013aa,Kelly:2013cpa}.
Furthermore, the inter-quark potential in charmonium was calculated 
using the HAL-QCD method, originally developed for the study of
the nucleon--nucleon potential \cite{Evans:2013yva}. 
At low temperatures, we observe agreement with the Cornell potential, and the potential flattens (weakens), as expected,  when the temperature increases.  
This is the first ab initio calculation of force between relativistic quarks as a function of temperature. The results are consistent with the expectation
that charmonium melts at high temperature. 

\begin{figure}
\centering
\includegraphics[width=14cm,clip]{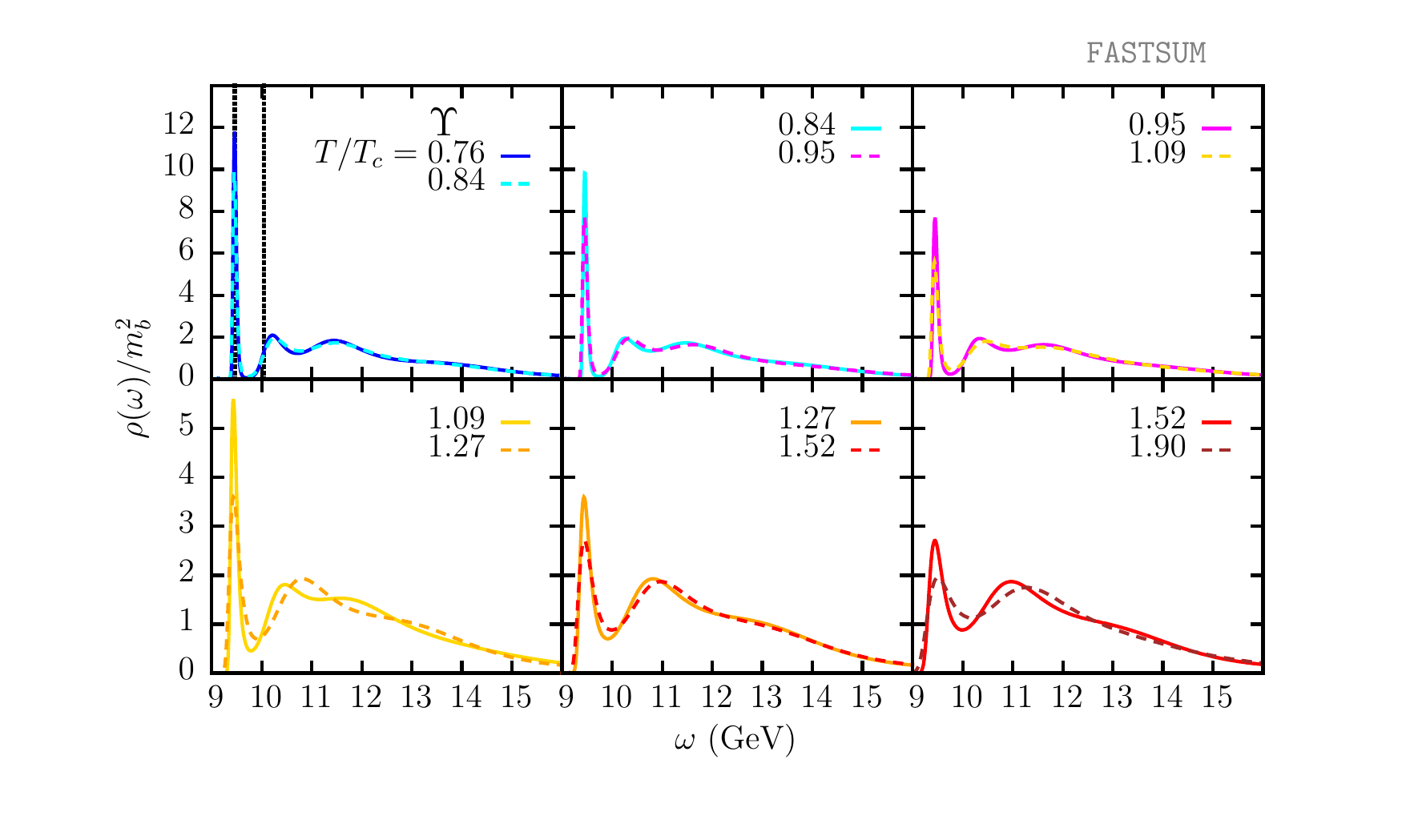}
\vskip -1 truecm
\caption{The spectral functions for the $\Upsilon$ at different
  temperatures, obtained using the maximum entropy method.}
\label{fig:u}       
\end{figure}

Bottomonium mesons have been studied using the 
NRQCD approximation for the bottom quark \cite{Aarts:2010ek}. 
We defer the discussion of
details to the next section, and here we focus on the main
results ---
the spectral functions. Note that in this case the low frequency limit
is excluded: transport peaks are sensitive to long-distance,
nearly constant modes and do not develop when winding along the Euclidean
time direction is suppressed. 
The results~\cite{Aarts:2014cda} for the Upsilon shown in fig.~\ref{fig:u}
clearly demonstrate the persistence of
the fundamental state  above $T_c$ as well as the suppression
of the excited states. These patterns should be contrasted, for instance,
with the one observed by the CMS experiment: for an estimated temperature
of about $420$ MeV the excited peaks of the invariant mass distribution
are suppressed. Consider now the rate of production of muon pairs
$\frac{dN_{\mu \bar \mu} }{d^4xd^4q} = F(q,T,...)\rho(\omega)$.
The connection between the invariant mass distribution 
and the spectral function is clear, although the dynamical
factor $F$ is largely unknown. Understanding in detail this connection
is an important aspect of ongoing research \cite{Antinori:2014xma}.
In the following we will limit ourselves to the presentation and discussion
of our spectral functions.  
The comparison of important features of our results --- masses, as seen
in the central peak positions, and associated widths --- with effective models
is satisfactory, and gives us further confidence in our analysis.  

\begin{figure}
\centering
\includegraphics[width=14cm,clip]{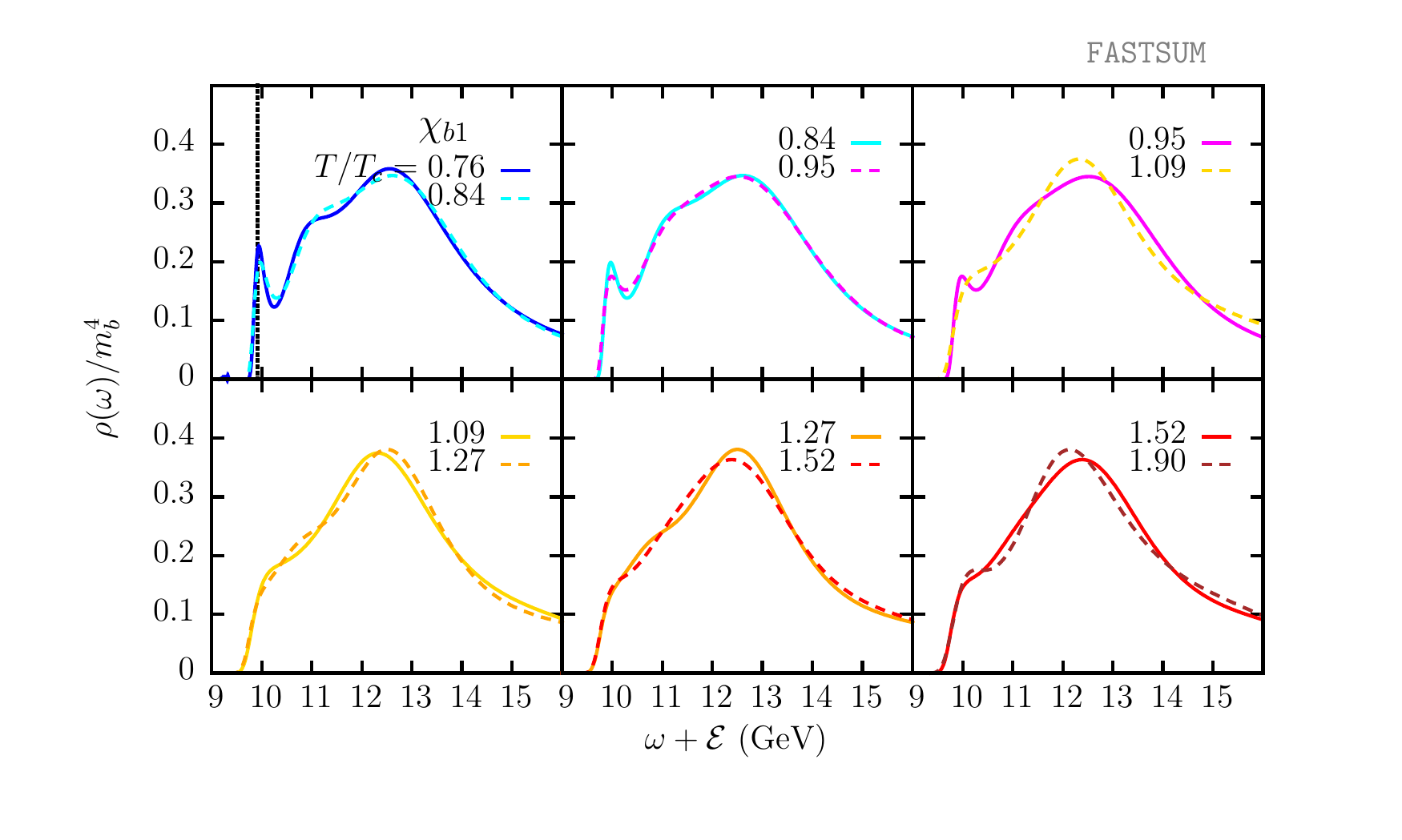}
\vskip -1 truecm
\caption{The spectral functions for the $\chi_b$ at different
  temperatures, obtained using the maximum entropy method.}
\label{fig:cp}      
\end{figure}

Our results for the $P$-wave $\chi_{b1}$ \cite{Aarts:2014cda} are shown in
fig.~\ref{fig:cp}. Here  
checks of the systematic errors are still in progress, and in particular
we would like to assess the fate of the fundamental state at $T_c$, possibly
before experimental results --- which are still lacking in this sector --- appear!

\section{Bottomonium in the plasma --- some details on our analysis}
We discuss here in more detail our results for the Upsilon and the $\chi_b$. 
The analysis starts with the computation of the correlators in Euclidean
time within the NRQCD formalism, which in turn are input to the spectral
functions presented above.

NRQCD is an effective field theory with power counting in the heavy quark velocity in the bottomonium rest frame.
The heavy quark and anti-quark fields decouple and their numbers 
are separately conserved.
Their propagators, $S(x)$, solve an initial-value problem whose discretization leads to the following choice for the evolution equation 
\begin{align}
    S(x+a_\tau\bm e_\tau)=
        \left(1-\frac{a_\tau H_0|_{\tau+a_\tau}}{2k}\right)^kU_\tau^\dagger(x)
        \left(1-\frac{a_\tau H_0|_{\tau}}{2k}\right)^k\left(1-{a_\tau\delta H}\right)
        S(x),
\end{align}
where $U_\tau(x)$ is the temporal gauge link at site $x$
and $\bm e_\tau$ the temporal unit vector. The leading order Hamiltonian is defined by
$H_0=-\frac{\Delta^{(2)}}{2m_b}$,
        with $\Delta^{(2n)}
        =\sum_{i=1}^3\left(\nabla_i^+\nabla_i^-\right)^{n} $.
The higher order covariant finite differences are written in terms of
the components of the usual forward and backward  first order
ones. Further details can be found e.g. in ref.~\cite{Aarts:2014cda}.
Only energy differences are physically significant in NRQCD because 
the rest-mass energy can be removed from the heavy quark dispersion relation 
by performing a field transformation. Since there 
is no rest mass term in the NRQCD action 
one  can dispense with the demanding constraint $ a \ll 1/m_b$.

In our most recent work~\cite{Aarts:2014cda}  we have tuned
the heavy quark mass by requiring the spin-averaged $1$S kinetic mass, $M_2(\overline{1\textrm{S}})=(M_2(\eta_b)+3M_2(\Upsilon))/4$, to be equal to its experimental value.
The tuned value of the heavy quark mass corresponds to $M_2(\overline{1\textrm S})=9560(110) \textrm{ MeV}$ which is consistent with the experimental value, $M_\mathrm{expt}(\overline{\mathrm{1S}})=9444.7(8)$ MeV. 

\begin{figure}[t]
    \centerline{
            \includegraphics[width=0.45\textwidth]{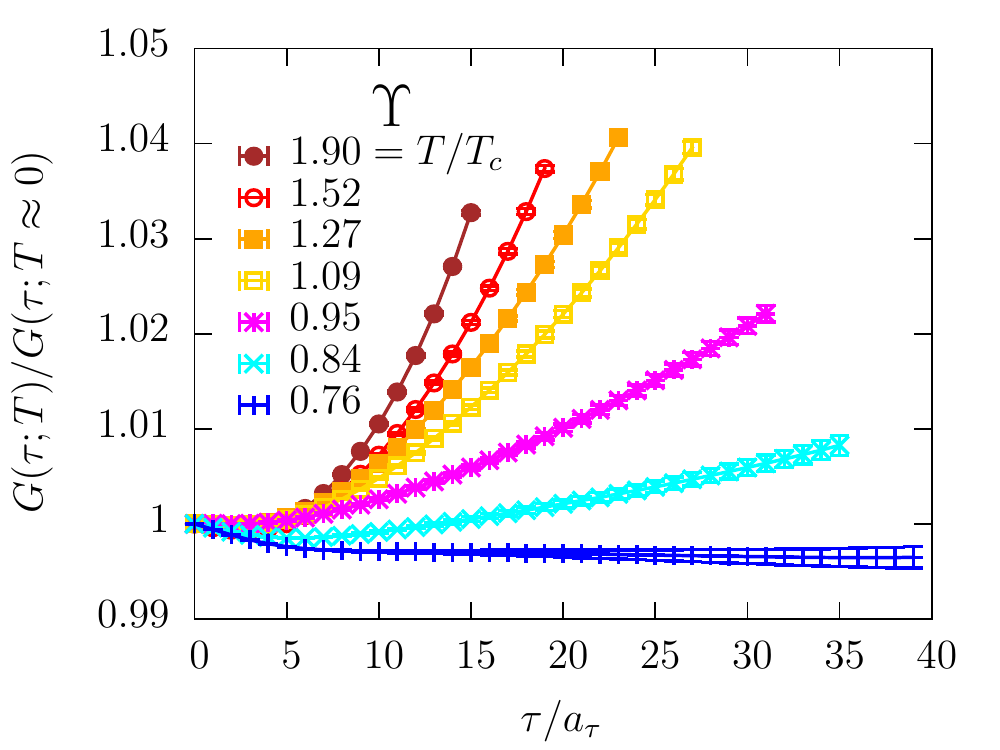}
            \includegraphics[width=0.45\textwidth]{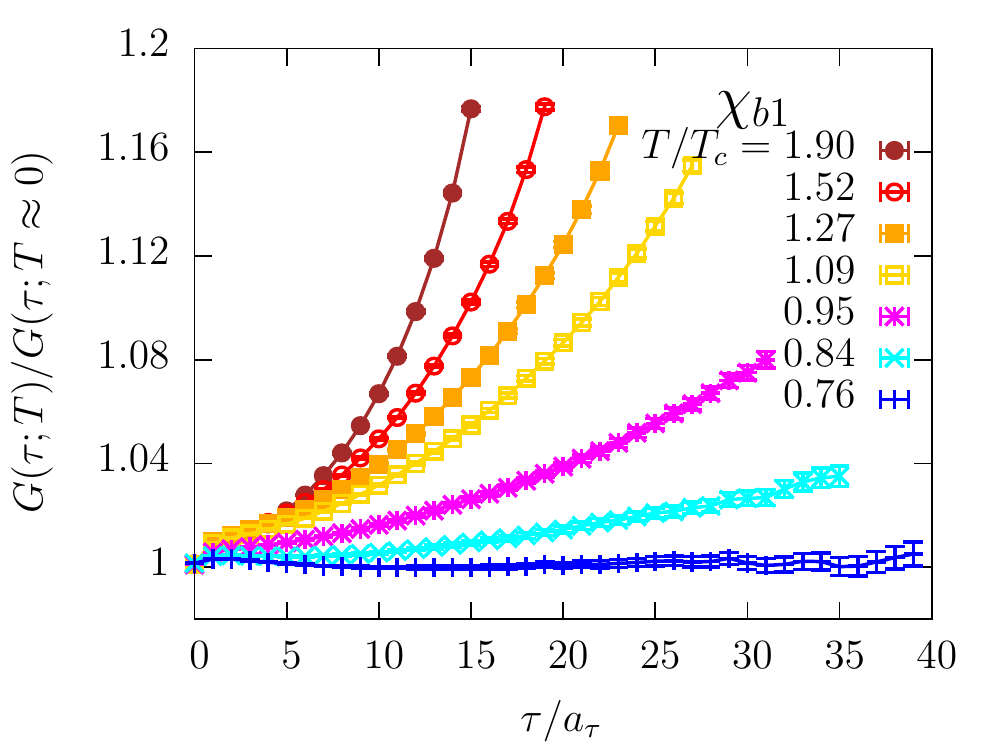}
    }
    \caption{Thermal modification, $G(\tau;T)/G(\tau;T\approx0)$, of the correlation functions in the $\Upsilon$ (left) and $\chi_{b1}$ (right) channels.}
    \label{fig:corrs_Tdep}
\end{figure}

\begin{figure}[t]
    \centerline{
              \includegraphics[width=0.45\textwidth]{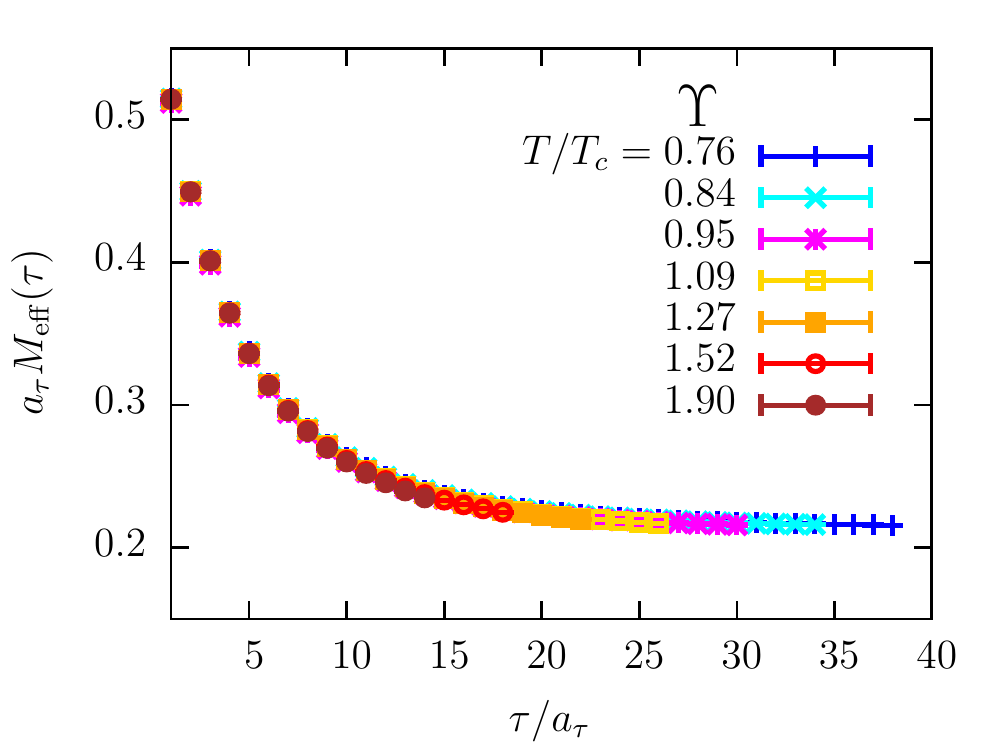}
              \includegraphics[width=0.45\textwidth]{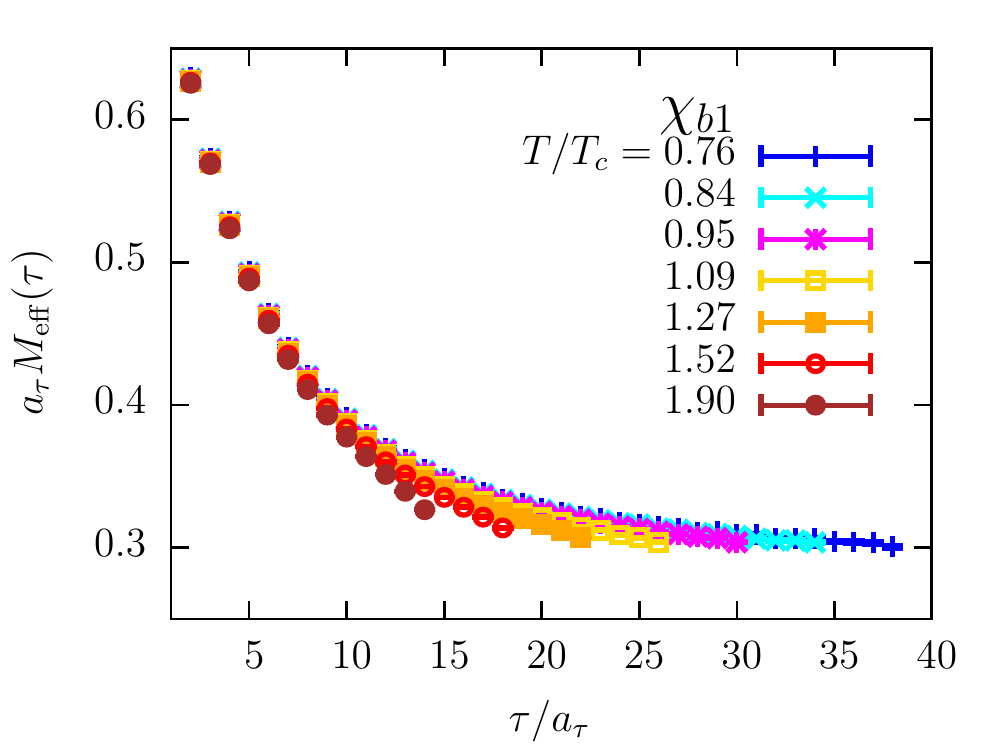}
    }
    \caption{Temperature dependence of the effective mass in the $\Upsilon$ (left) and the $\chi_{b1}$ (right) channels.}
    \label{fig:meff_Tdep}
\end{figure}

We now turn to the analysis of correlators. Consider first the infinite
temperature limit, i.e. the 
limiting case of free heavy quarks:
in continuum NRQCD the spectral functions are known~\cite{Burnier:2007qm}, and are given by\footnote{We have included a threshold, $\omega_0$, to account for the additive shift in the quarkonium energies. 
For free quarks the threshold occurs at $2m_b$, which within NRQCD corresponds to $\omega_0=0$.}
\begin{align}
    \rho_{\textrm{free}}(\omega)\propto(\omega-\omega_0)^{\alpha}\,\Theta(\omega-\omega_0),
        \quad\textrm{where}\quad
        \alpha=
        \begin{cases}
            1/2, &\textrm{S~wave};\\
            3/2, &\textrm{P~wave}.
        \end{cases}
    \label{eq:rhofree}
\end{align}
The correlation functions then have the following behaviour
\begin{align}
    G_{\mathrm{free}}(\tau)\propto \frac{e^{-\omega_0\tau}}{\tau^{\alpha+1}}.
    \label{eq:powerlaw}
\end{align}
To show their temperature
dependence we consider 
the ratios of the correlation functions at finite temperature to those
at zero temperature. They are shown in
fig.~\ref{fig:corrs_Tdep}. We see that the thermal
  modifications are much larger in the P-wave than in the S-wave channel.
A useful numerical tool is the so-called effective mass $M_\mathrm{eff}(\tau)$:
\begin{align}
    M_\mathrm{eff}(\tau) \equiv -\frac{1}{G(\tau)}\frac{dG(\tau)}{d\tau}
    \quad\stackrel{G=G_\mathrm{free}}{\longrightarrow}\quad
    \omega_0+\frac{\alpha+1}{\tau}.
    \label{eq:effmass}
\end{align}
The results are shown in fig.~\ref{fig:meff_Tdep}. 
The S-wave effective mass displays little temperature dependence
(left) but a clear effect is seen in the P-wave channel
effective mass (right).
In ref.~\cite{Aarts:2010ek} it was also observed that the S-wave effective mass showed little variation with temperature while the temperature dependence in the P-wave channel effective mass was even more pronounced than visible here.
 
\begin{figure}[t]
    \centerline{
            \includegraphics[width=0.45\textwidth]{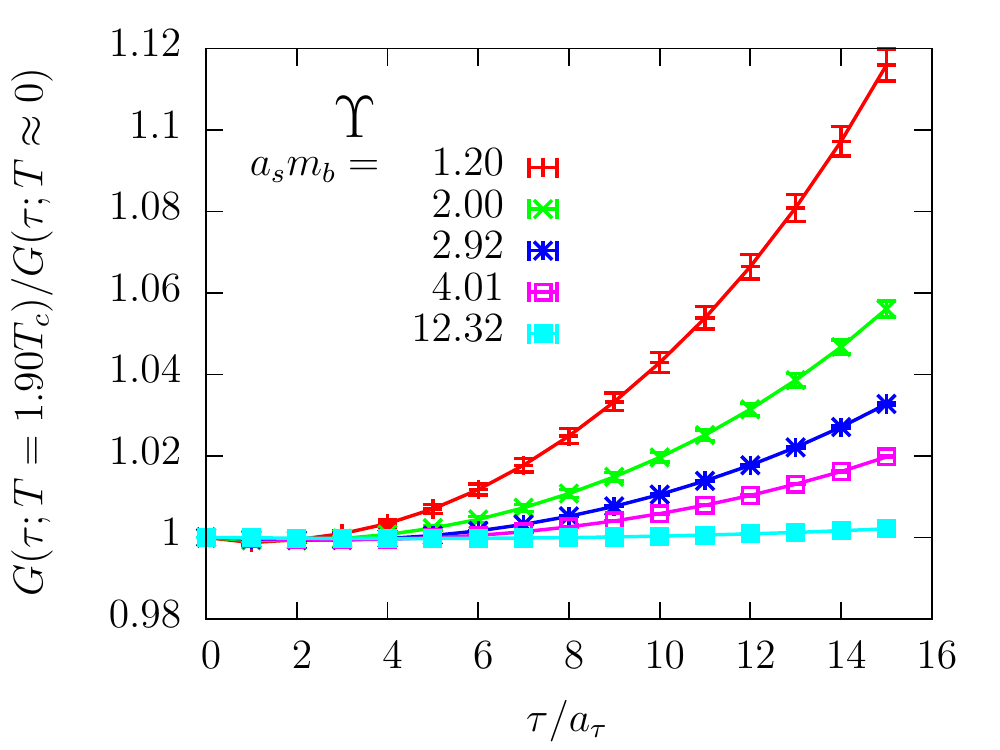}%
            \includegraphics[width=0.45\textwidth]{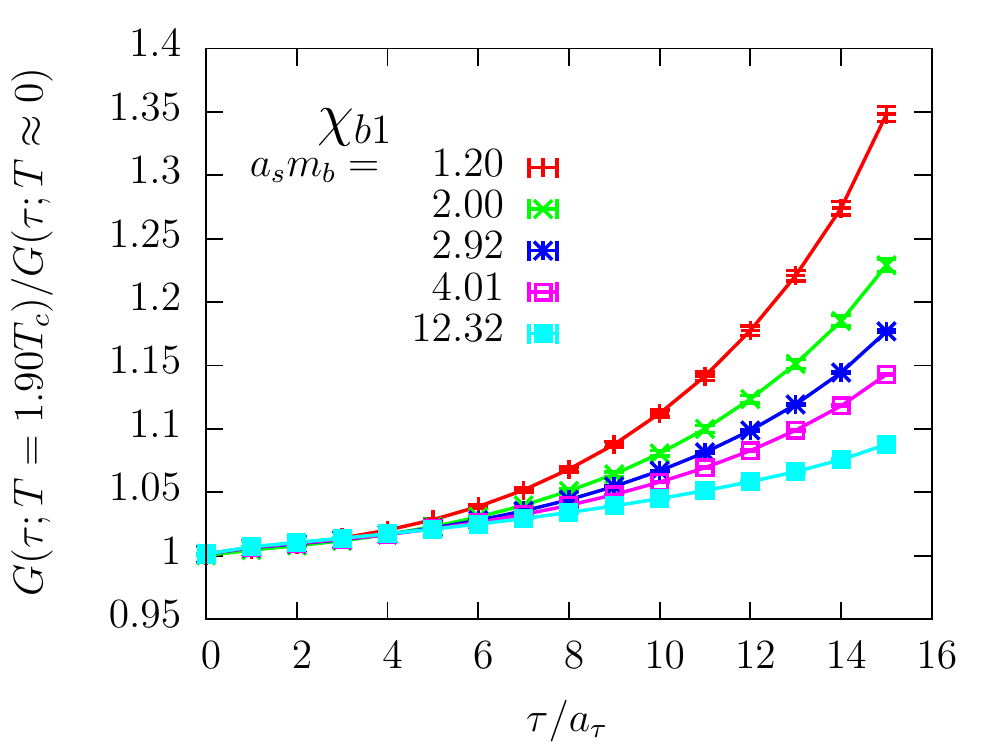}%
    }
    \caption{Dependence on the heavy quark mass of the modification in the correlators at the highest accessible temperature, $T/T_c=1.90$, in the $\Upsilon$ (left) and $\chi_{b1}$ (right) channels.}
    \label{fig:mbdep}
\end{figure}
These results clearly show a temperature dependence, but it is not easy to
assess with confidence the fate of bound states. While in real time the information 
on the long term dynamics is fully accessible,  in imaginary time all the
information is squeezed within the periodicity $\tau_P = 1/T$. One would
need an extremely high accuracy on extremely fine lattices to make quantitative
statements from the correlators alone. 
This further motivates an analysis in terms of spectral functions.

\section{Discussion}
In summary, we have a coherent scenario for the $\Upsilon$:
the fundamental state survives up to at least twice the critical temperature,
while the excited states dissolve. With the caveats mentioned above, this 
is consistent with the observations of
CMS, ALICE and PHENIX. The fundamental state has some modifications
whose basic features can be captured by effective field
theories. However, at a temperature of about $420$ MeV ALICE results \cite{Das}
indicate that the suppression of $\Upsilon$ and $J/\psi$ as a function
of the number of participants is comparable, within the present uncertainties.
This can be explained by the $J/\psi$ being more suppressed, but also
more sensitive to regeneration, the two effects competing in such a way
that the resulting $R_{AA}$ is similar to that of the $\Upsilon$. When
comparing RHIC and LHC results, it is found that the nuclear modification
factor at RHIC is smaller than at the LHC --- the so called quarkonium 
suppression puzzle.   New theoretical ideas have been put forward to
interpret this behaviour \cite{Kharzeev:2014pha}. All  
 this confirms the interest in ab initio
lattice studies of charmonia and bottomonia in hot matter with full
control of systematical errors. On the lattice we might also take advantage
of the freedom to simulate arbitrary masses: some preliminary 
results were presented in ref.~\cite{Kim:2012by}  and the most recent ones 
\cite{Aarts:2014cda} are shown in fig.~\ref{fig:mbdep}. Guided by these analysis  we might be
able   to locate a melting line in the temperature--mass plane which 
passes through the individual melting temperatures observed in
different  channels.  These studies might help unravel general features of
the dissolutions of heavy states and their interrelation with gauge dynamics. 

One important next step is a full control over matter content in our lattice
simulations. The simulations reported here have been performed with
$\frac{m_\pi}{m_\rho} \simeq 0.4$, and with $m_s$ either
set to infinity or to its physical value.  We aim at physical $m_{u,d,s}$ masses
which should correspond to the correct matter
content in the range $T \le 400$ MeV. Above $400$ MeV a dynamical
charm quark might become relevant as well. 

We have already mentioned the subtleties related with the reconstruction
of the spectral functions.  To gain confidence in our analysis we
will continue  
cross checking MEM results with those based on the novel Bayesian
approach \cite{Burnier:2013nla,Lat,Conf}; applications of a  generalised integral transform
might ease the inversion task \cite{Pederiva:2014qea}; and
model calculations will provide very useful testbeds for these new techiques
\cite{Colangelo:2012jy,Giannuzzi:2014rha}.  

\vskip 1 truecm
\noindent{\bf Acknowledgements} 
It is a pleasure to thank Yannis Burnier and Alexander Rothkopf
for many useful discussions. 
MpL wishes to thank the organisers of QCD@Work 2014 
for their very nice hospitality and a most interesting meeting.  
AK acknowledges financial support through the Irish Research Council.

\end{document}